\documentclass[pra,twocolumn,nofootinbib,aps,amsmath,showkeys,showpacs,floatfix]{revtex4-1}

\usepackage{graphicx}
\usepackage{dcolumn}
\usepackage{bm}
\usepackage{color}

\newcommand{\bs}{\boldsymbol}
\newcommand{\beq}{\begin{equation}}
\newcommand{\bk}[2]{\left\langle\,#1\left|\,#2\,\right\rangle\right.} 
\newcommand{\ket}[1]{\left|#1\right\rangle} 
\newcommand{\eeq}{\end{equation}}
\newcommand{\bpm}{\begin{pmatrix}}
\newcommand{\epm}{\end{pmatrix}}
\newcommand{\bea}{\begin{eqnarray}}
\newcommand{\eea}{\end{eqnarray}}

\newcommand{\ignore}[1]{}

\begin{document}


\title{Strongly correlated states of a small cold atomic cloud from 
geometric gauge fields}

\author{B. Juli\'a-D\'{\i}az$^{1,3}$, D. Dagnino$^1$, K. J. G\"{u}nter$^2$,
  T. Gra\ss$^3$,  N. Barber\'an$^1$, M. Lewenstein$^{3,4}$, and J. Dalibard$^2$}

\affiliation{$^1$Dept. ECM, Facultat de F\'isica, U. Barcelona, 08028 Barcelona, Spain}
\affiliation{$^2$Laboratoire Kastler Brossel, CNRS, UPMC, Ecole Normale 
Sup\'erieure, 24 rue Lhomond, 75005 Paris, France}
\affiliation{$^3$ICFO-Institut de Ci\`encies Fot\`oniques, Parc Mediterrani 
de la Tecnologia, 08860 Barcelona Spain}
\affiliation{$^4$ICREA-Instituci\'o Catalana de Recerca i Estudis Avan\c cats, 
08010 Barcelona, Spain}

\vskip11mm

\medskip

\begin{abstract}
Using exact diagonalization for a small system of cold bosonic atoms, we
analyze the emergence of strongly correlated states in the presence of 
an artificial magnetic field. This gauge field is generated by a laser beam 
that couples two internal atomic states, and  it is related to Berry's 
geometrical phase that emerges when an atom follows adiabatically one of 
the two eigenstates of the atom--laser coupling. Our approach allows us 
to go beyond the adiabatic  approximation, and to characterize the 
generalized Laughlin wave functions that appear in the strong magnetic 
field limit.  
\end{abstract}

\vskip2mm

\pacs{73.43.-f,03.65.Vf,37.10.Vz}
\keywords{Gauge fields. Ultracold Bose gases. Correlated states. Exact diagonalization.}
\maketitle
\vfill
\eject

\vskip11mm

\medskip

\section{Introduction}
\label{s1}

Trapped atomic gases provide a unique playground to address many-body 
quantum physics in a very controlled way~\cite{Lewenstein:2007,Bloch:2008}. 
Studies of ultracold atoms submitted to artificial gauge fields are 
particularly interesting in this context, since they establish a link 
with the physics of the quantum Hall effect. Especially intriguing is 
the possibility of realizing strongly correlated states of the gas, 
such as an atomic analog of the celebrated Laughlin state~\cite{Cooper:2008}.

One way to simulate orbital magnetism is to rotate the trapped 
gas~\cite{Cooper:2008,Fetter:2009}. One uses in this case the 
analogy between the Coriolis force that appears in the rotating 
frame and the Lorentz force acting on a charged particle in a 
magnetic field. This technique allows one to nucleate vortices 
and observe their ordering in an Abrikosov lattice~\cite{Fetter:2009}. 
Another promising method takes advantage of Berry's geometrical 
phase that appears when a moving atom with multiple internal 
levels follows adiabatically a non-trivial linear combination 
of these levels~\cite{Berry:1984}. This can be achieved in 
practice by illuminating the gas with laser beams that induce 
a spatially varying coupling between atomic internal levels~\cite{Dum:1996} 
(for a review of recent proposals, see e.g.~\cite{Dalibard:2010}). 
Recently spectacular experimental progress has been made with this 
technique, leading here also to the observation of quantized vortices~\cite{Lin:2009b}.

In this article we focus on the generation of strongly correlated 
states of the atomic gas with geometrical gauge fields. We show 
how Laughlin-type states emerge in a small quasi two-dimensional 
system of trapped bosonic atoms when two internal states are 
coupled by a spatially varying laser field. The key point of our 
approach is to go beyond the adiabatic approximation and study 
how the possibility of transitions between the internal states 
modifies the external ground state of the gas. We perform exact 
diagonalization for $N=4$ particles, in order to analyze the 
overlap between the exact ground state of the system and the 
Laughlin wave function, as a function of the strength of the 
atom--laser  coupling. We identify a region of parameter space 
in which the ground state, despite having a small overlap with 
the exact Laughlin state, has an interaction energy close to 
zero, a large angular momentum, and a large entropy. 
We show that it can be represented as a Laughlin-like state 
with modified Jastrow factor. 

To better frame our work, let us first emphasize that it 
is well known that the adiabatic accumulation of Berry phase 
induces artificial gauge fields in electroneutral quantum 
systems~\cite{Berry:1984,wen}. These artificial gauge 
fields of geometric origin are nowadays considered as an 
important new framework in which strongly correlated quantum 
states related to fractional quantum hall physics can be 
engineered in systems of ultracold atoms~\cite{Dalibard:2010}. 
The major contribution of the present article is to go beyond 
adiabaticity to show that 
the main fingerprints of the strongly correlated quantum 
state are preserved in a significantly broad region of the 
experimental parameter space, as demonstrated by detailed 
numerical calculations. This objective is discussed 
in detail in the framework of the physics of few-body systems, 
independently of their attainability in the thermodynamic 
limit, which is beyond the scope of the present article. This 
goal is already a realistic one as there are nowadays a number 
of experimental groups able of dealing with small bosonic 
clouds using several techniques~\cite{kino,gemelke}. These 
experimental developments have triggered a number of theoretical 
proposals focusing on the production of strongly correlated 
quantum states in small atomic clouds~\cite{popp,ronca,ronca2}.

The article is organized in the following way. First, in Sec.~\ref{s2}
we present the scheme used to generate the artificial gauge field 
together with the formalism employed. In Sec.~\ref{s3} we present 
our results, discussing in detail the properties of the Laughlin-like 
strongly correlated states appearing for different values of the 
external control parameters. In Sec.~\ref{ss6} we study the analytical 
representation of the ground state in the strongly correlated
Laughlin-like region. Finally, in Sec.~\ref{s4} we provide 
some conclusions which can be extracted from our work.

\section{Theoretical model}
\label{s2}

\begin{figure*}[t]
\begin{center}
\includegraphics*[width=120mm]{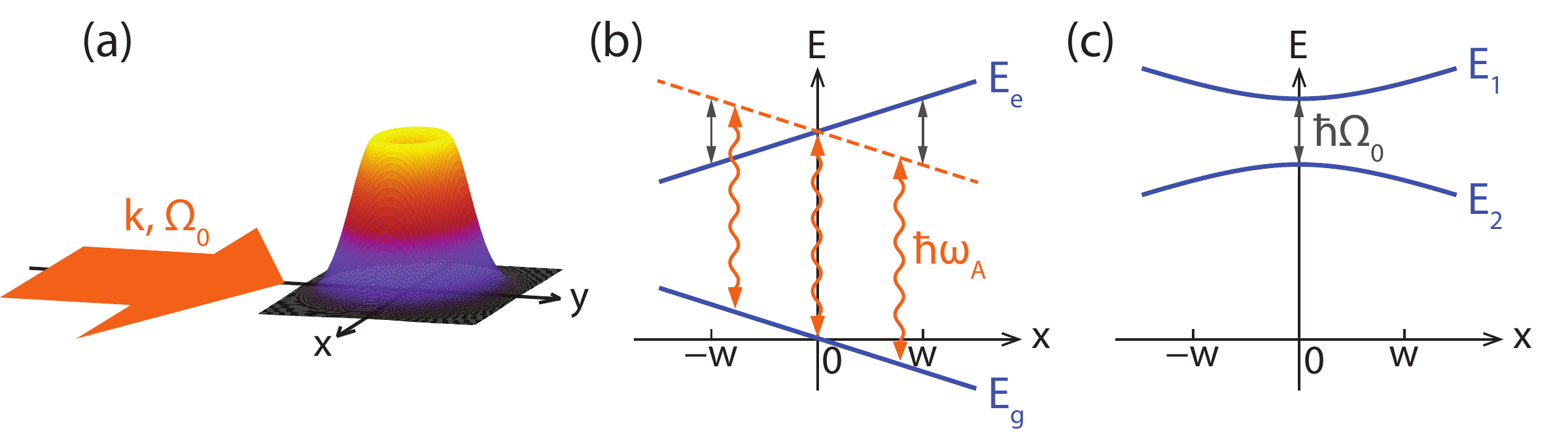}
\caption{(Color online) Schematic illustration of the 
considered setup. (a) Atoms are trapped in the $x$--$y$ 
plane and illuminated with a plane wave propagating along 
the $y$ direction. (b) The energy difference between the 
two internal states that are coupled by the laser field 
varies linearly along the $x$ direction. (c) Energy eigenvalues 
of the atom--laser coupling in the rotating wave approximation.
\label{fig0}}
\end{center}
\end{figure*}
We consider a small quasi two-dimensional ensemble of 
harmonically trapped bosonic atoms in the $x$--$y$ plane 
interacting with a single laser field treated in a classical 
way. The single-particle Hamiltonian is given by
\beq
\hat H_{\rm sp}=
\frac{\bs p^2}{2M}
+ V(\bs r)
+ \hat{H}_{\rm AL} \,,
\label{eqsp} 
\eeq 
where $M$ is the atomic mass and $V(\bs r)$ is an external potential 
confining the atoms in the plane. $\hat{H}_{\rm AL}$ includes the 
atom--laser coupling as well as the internal energies. In order to 
minimize the technical aspects of the proposal, we consider here a 
very simple laser configuration to generate the geometrical gauge 
field. However, it is straightforward to generalize our method to 
more complex situations. The laser field is a plane wave with wave 
number $k$ and frequency $\omega_{\rm L}$ propagating along the 
direction $y$ [see Fig.~\ref{fig0}(a)]. It couples two internal 
atomic states $|{\rm g}\rangle$ and $|{\rm e}\rangle$ with a 
strength that is given by the Rabi frequency $\Omega_0$. 

The atom-laser term comes from the coupling of the 
electric dipole of the atom with the electric field of the laser. It 
can be written as,  
\bea
\hat{H}_{\rm AL}&=& 
E_{\rm g} |{\rm g}\rangle \langle {\rm g}| 
+ E_{\rm e} |{\rm e}\rangle \langle {\rm e}| \nonumber \\
&+& {\hbar \Omega_0} \cos( \omega_{\rm L} t - \phi) 
\left(|{\rm e}\rangle \langle {\rm g}| + |{\rm g}\rangle \langle {\rm e}|\right)\,,
\eea
with $\phi=k y$. We neglect the spontaneous emission rate of photons 
from the excited state $|{\rm e}\rangle$, which is a realistic 
assumption if the intercombination line of alkali-earth or 
Ytterbium atoms is used. For instance, for Ytterbium atoms 
the state $|{\rm e}\rangle$ could be taken as the first 
excited level $6 ^3P_0$ which has a very long lifetime 
of $\sim 10$~s~\cite{yb}. Importantly, we further assume that 
the energies $E_{\rm g} = -\hbar\Omega_0 x/(2w)$ and 
$E_{\rm e} = \hbar\omega_{\rm A} + \hbar\Omega_0 x/(2w)$ of the 
uncoupled internal states vary linearly along $x$ in a length 
scale set by the parameter $w$, as sketched in Fig.~\ref{fig0}(b). 
This can be achieved experimentally either 
by profiting the Zeeman effect, i.e. applying a real 
magnetic field gradient to the system, or by using the a.c. 
Stark shift produced by an extra laser beam with an intensity gradient.  
We suppose that the laser is resonant with the atoms in 
$x=0$, i.e. $\omega_{\rm L}=\omega_{\rm A}$. 
Using the rotating-wave approximation, $\hat{H}_{\rm AL}$ can 
be written in the frame 
rotating with the laser frequency $\omega_L$ and in the 
 $\{ |{\rm e}\rangle,\, |{\rm g}\rangle \}$ basis~\cite{cohe92} as, 
\beq
\hat{H}_{\rm AL} =\frac{\hbar \Omega}{2}
\begin{pmatrix}
\cos\theta & {\rm e}^{i\phi} \sin\theta\\
{\rm e}^{-i\phi} \sin\theta & -\cos\theta
\end{pmatrix} \, ,
\label{eqal} 
\eeq 
where $\Omega=\Omega_0\sqrt{1+x^2/w^2}$, and $\tan \theta=w/x$.

It is convenient to rewrite Eq.~(\ref{eqsp}) in the basis of 
the local eigenvectors of $\hat{H}_\mathrm{AL}$,
$|\psi_1\rangle_{\bf r}$ and $|\psi_2\rangle_{\bf r}$, associated to the 
eigenvalues $\hbar \Omega/2$ and $-\hbar \Omega/2$, respectively. 
These can be written in the $\{\ket{e},\ket{g}\}$ basis as, 
\bea
\ket{\psi_1}_{\bf r} &=&{\rm e}^{-iG}
\begin{pmatrix}
\cos{\theta/2}\,\, {\rm e}^{i\phi/2}\\
\sin{\theta/2}\,\,{\rm e}^{-i\phi/2} 
\end{pmatrix} \nonumber \\
\ket{\psi_2}_{\bf r} &=&{\rm e}^{iG}
\begin{pmatrix}
-\sin{\theta/2}\,\, {\rm e}^{i\phi/2}\\
\cos{\theta/2}\,\,{\rm e}^{-i\phi/2} 
\end{pmatrix} \, ,
\label{eq4}
\eea
where $G=\frac{kxy}{4w}$. This particular form 
of Eq.~(\ref{eq4}) allows us in what follows 
to obtain a fully symmetric $H_{22}$, see 
Eq.~(\ref{eq9}). Let us emphasize that the choice 
of the phase factor in front of these eigenstates is 
nothing but a gauge choice for the following.

The atomic state can be then expressed as
\begin{equation}
\chi(\bs r,t) = 
a_1(\bs r,t)\otimes |\psi_1\rangle_{\bf r} 
+ a_2(\bs r,t)\otimes |\psi_2\rangle_{\bf r}\, ,
\label{eq7}
\end{equation}
where $a_i$ captures the dynamics of the center of mass and 
$|\psi_i\rangle_{\bf r}$ of the internal degree of 
freedom (in the following we drop the subindex ${\bs r}$ 
in the kets $\ket{\psi_{1,2}}$ to simplify the notation). 
Projecting onto the basis $\{|\psi_i\rangle \}$, and noting that 
\beq
\nabla_{\bs r}\,( a_j \psi_j) = 
a_j (\nabla_{\bs r}\, \psi_j) + (\nabla_{\bs r} \,a_j) \psi_j \,,
\eeq
the single-particle Hamiltonian, Eq.~(\ref{eqsp}), is represented 
by the $2\times 2$ matrix $\hat H_{\rm sp}=\left[H_{ij} \right]$ 
acting on the spinor $\left[ a_1 (\bs r,t), a_2(\bs r,t)  \right]$. 
We find in particular~\cite{Mead:1979,Berry:1989}
\beq
H_{jj}=\frac{\left[\bs p-\epsilon_j \bs A \right]^2}{2M} + U+V +\epsilon_j\frac{\hbar \Omega}{2}\,,
\label{eq14} 
\eeq 
with $\epsilon_1=1$ and $\epsilon_2=-1$, where we have defined 
\beq
\bs A(\bs r)=-i\hbar \bk{\psi_{2}}{\bs \nabla_{\bs r}\psi_{2}}
\eeq
and
\beq
U(\bs r)=\frac{\hbar^2}{2M}[\bk{\bs \nabla_{\bs r}\psi_{2}}{\bs \nabla_{\bs r}\psi_{2}}+(\bk{\psi_{2}}{\bs \nabla_{\bs r}\psi_{2}})^2]\,.
\eeq
For the chosen gauge they read, 
\bea
\bs A(\bs r)&=&
\hbar k \left[ \frac{y}{4w} , \frac{x}{4w}-\frac{x}{2\sqrt{x^2+w^2}}
\right]\,,
\label{eq12} \\
 U(\bs r)&=&
\frac{\hbar^2 w^2}{8M\left(x^2+w^2\right)}\left(k^2+\frac{1}{x^2+w^2}\right) \,.
\eea

We consider atomic clouds extending over distances smaller
than $w$. This allows us to expand the matrix elements $H_{ij}$ 
up to second order in $x$ and $y$. In this approximation, we 
recover the symmetric gauge expression 
$\bs A(\bs r)=\frac{\hbar  k}{4w}(y,-x)$ and the 
artificial magnetic field reduces to 
$\bs B_j = \epsilon_j \hbar k/(2w)\, \hat{\bs z}$ for an 
atom in $|\psi_j\rangle$. The specific choice of the phase 
factors in Eq.~(\ref{eq4}) is in fact obtained by imposing 
as a constraint the symmetric gauge at this step. 
Finally, we fix the external potential $V(\bs r)$ such that 
the total confinement for the spinor component $a_2$ 
is isotropic with frequency $\omega_\perp$:
\beq
\frac{1}{2}m\omega_\perp^2 (x^2+y^2)=U(\bs r)+V(\bs r) 
-\frac{\hbar \Omega(\bs r)}{2}+\frac{\bs A^2(\bs r)}{2M}\,.
\eeq
The Hamiltonian 
\bea
H_{22}&=&p^2/2M +\bs p\cdot\bs A/M +M\omega_\perp^2 r^2/2
\nonumber
\\
& = & \frac{({\bs p} + {\bs A})^2}{2M}
+\frac{M \omega_\perp^2}{2}(1-\eta^2) r^2
\label{eq9}
\eea
is thus circularly symmetric and its eigenfunctions are the 
Fock-Darwin (FD) functions $\phi_{\ell, n}$, with $\ell$ and 
$n$ denoting the single-particle angular momentum and the 
Landau level, respectively. The magnetic field 
strength is characterized by $\eta\equiv \omega_{\rm  c}/2\omega_\perp$, 
with $\omega_{\rm c}=|B_j|/M=\hbar k/ (2Mw)$ the `cyclotron frequency'. 

The interesting regime for addressing quantum Hall physics 
corresponds to quasi-flat Landau levels, which occurs when 
the magnetic field strength $\eta$ is comparable to 1. 
The energies of the states of the Lowest Landau Level (LLL), $n=0$, 
are 
$ E_{\ell,0}= \hbar \omega_\perp \left[1+  \ell (1-\eta) + (k^2\lambda^2_{\perp}/ 8) +
\lambda^2_{\perp}/(8 w^2)\right]$, where $\lambda_\perp=\sqrt{\hbar /M  \omega_\perp}$. 

Relevant energy scales of the single-particle problem are $\hbar\Omega_0$, 
which characterizes the internal atomic dynamics, and the  recoil 
energy $E_{\rm R}=\hbar^2 k^2/(2M)$, which gives the scale for the
 kinetic energy of the atomic center-of-mass motion when it absorbs 
or emits a single photon. For $\hbar\Omega_0 \gg E_{\rm R}$ the adiabatic 
approximation holds and the atoms initially prepared in the internal 
state $|\psi_2\rangle$ will remain in this state in the course of 
their evolution \cite{Dalibard:2010}. The single-particle Hamiltonian 
$H_{22}$, in combination with repulsive contact interactions, then 
leads to quantum Hall-like physics which has already been extensively 
studied \cite{Cooper:2008}. Our goal here is to consider corrections 
to the adiabatic approximation and to analyze in which respect these 
corrections still allow one to reach strongly correlated states. 
This aspect is particularly important from an experimental point 
of view, since the accessible range of $\Omega_0$ is limited if 
one wants to avoid undesired excitation of atoms in the sample 
to higher levels and/or an unwanted laser assisted modification of 
the atom-atom interaction. Note that the strength of the atom--laser 
coupling, characterized by $\Omega_0$, is distinct from the strength of the 
magnetic field, characterized by $\eta$. Because the magnetic 
field has a geometric origin, $\eta$ is independent of the 
atom--laser coupling as long as the adiabatic approximation is meaningful.

In the following we consider the situation where $\hbar\Omega_0$ is still
relatively large compared to $E_{\rm R}$, so that we can treat the coupling
between the internal subspaces related to $|\psi_{1,2}\rangle$ in a
perturbative manner. In a systematic expansion in powers of $\Omega_0^{-1}$,
the first correction to the adiabatic approximation consists (for the spinor
component $a_2$) in replacing $H_{22}$ by the effective Hamiltonian~\cite{cohe92}
\beq
H_{22}^{\rm eff} =  H_{22}  - {H_{21} H_{12}\over \hbar \Omega_0}\,.
\eeq 
The additional term $H_{21}H_{12}/(\hbar\Omega_0)$, which does not 
commute with the total angular momentum, is somewhat reminiscent of the 
anisotropic potential that is applied to set an atomic cloud in 
rotation \cite{Parke:2008,Dagnino:2009}. It is however mathematically 
more involved and physically richer, as it includes not only powers of 
$x$ and $y$, but also spatial derivatives with respect to these variables, 
see Appendix~\ref{ap1} for its explicit form.

\section{Ground state properties}
\label{s3}

The quasi-degeneracy in the LLL can lead to strong correlations 
as the interaction picks a many-body ground state for the 
system. The interaction between the atoms is well described 
by a contact interaction with a coupling constant 
$g=\sqrt{8\pi} (a_s/\l)$ for the quasi two-dimensional 
confinement. Here $a_s$ is the s-wave scattering length 
and $l$ the thickness of the gas in the strongly-confined 
$z$ direction. The many-body Hamiltonian then reads
\begin{equation}
H = \sum_{i=1}^N H_{22}^{\rm eff}(i) + \frac{\hbar^2 g}{M}\sum_{i<j}\delta(\bs r_i-\bs r_j) \,.
\label{eq26}
\end{equation}
Using an algorithm for exact diagonalization within the LLL 
of $H_{22}$, we have determined the many-body ground state 
(GS) of the system, providing phase diagrams of several relevant 
average values characterizing the system in a broad range of 
laser couplings, $\Omega_0$, and magnetic field strengths, $\eta$. 
To ensure the validity of the LLL assumption, we demand that the
difference in energy between different Landau levels is larger than the
kinetic energy of any particle in a FD state inside a Landau level. 
In addition, in the full many body problem the interaction energy per particle
is always much smaller than the energy difference between adjacent 
Landau levels. 
The main results are summarized in
Figs.~\ref{fig:summary1},~\ref{fig:summary2}, 
and~\ref{fig:summary3} and discussed in subsections \ref{ss1},~\ref{ss2} 
and~\ref{ss3}, respectively. In subsection~\ref{ss4} we analyze the internal 
correlations in the Laughlin state and in~\ref{ss5} we address the important 
problem of the role of excitations in the Laughlin like region.  
All the calculations are performed for $N=4$ atoms, 
$k=10/\lambda_\perp$, and $gN=6$. 
As the perturbation $H_{21} H_{12}$ breaks the rotational 
symmetry, we cannot carry out the exact diagonalization by 
restricting ourselves to a subspace with fixed angular momentum 
as in standard literature. To achieve convergent numerical 
results, we need to include a large number of $L$ subspaces, 
and this number grows as $\hbar\Omega_0/E_R$ is decreased. In 
all cases we consider all subspaces with $0\leq L <L_{\rm max}$, 
where $L_{\rm max}$ is chosen to ensure convergency. For 
definiteness let us quote the size of the Hilbert spaces 
considered in our work. For $N=4$ we require $L_{\rm max}=28$ 
for most of the numerical results reported in the paper, 
which results in a Hilbert space size of $2157$. 
This rapidly growing size of the Hilbert space as $N$ is 
increased, together with our explicit interest in 
providing fine-step phase diagrams varying both parameters, 
$\eta$ and $\hbar \Omega_0$ in a broad region makes our 
full study already computationally very extensive, i.e. 
the calculations presented in this article require 
on the order of 2 weeks on a single 2 GHz processor.

\subsection{Angular momentum}
\label{ss1}

\begin{figure}[t]
\begin{center}
\includegraphics*[width=90mm]{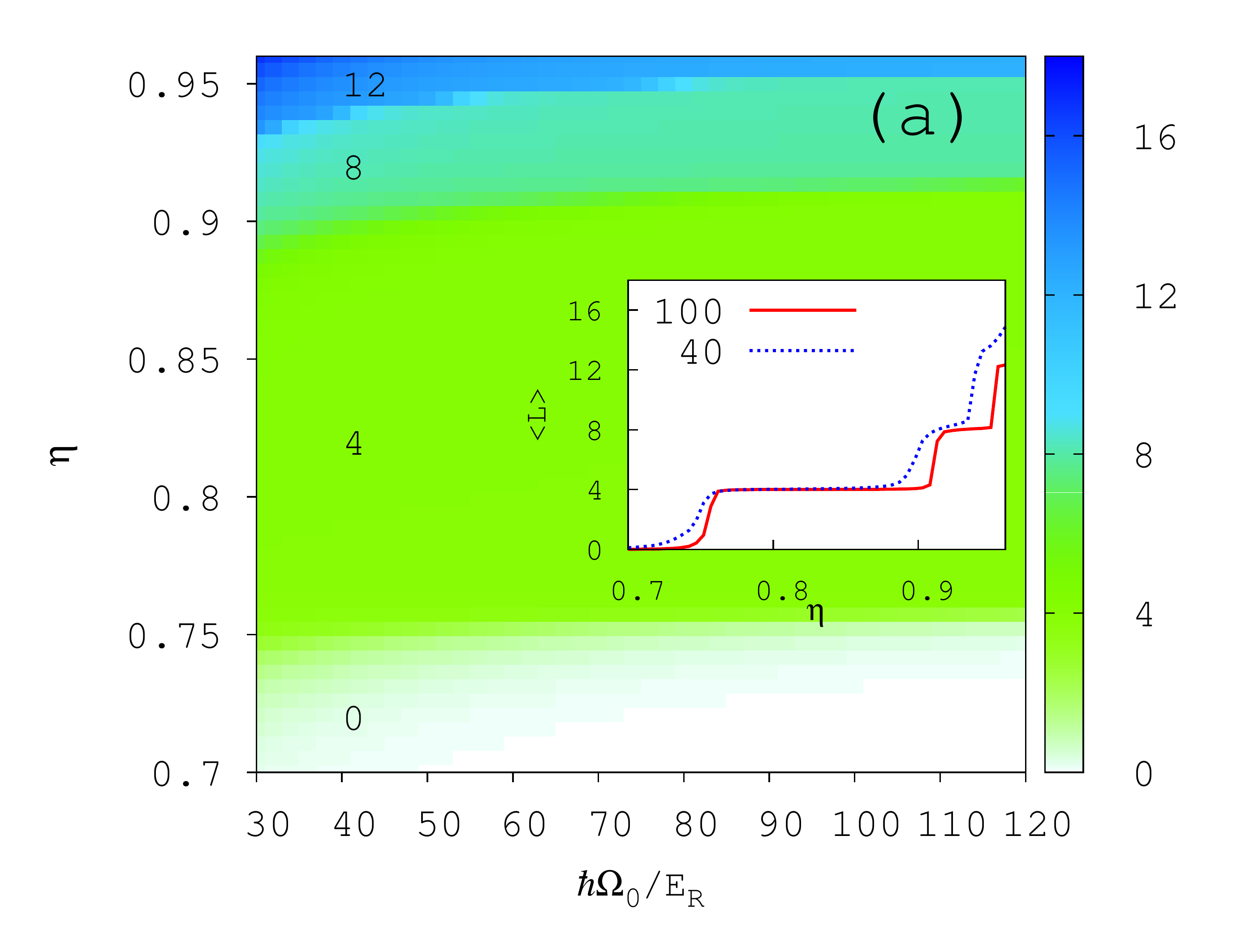}
\caption{(Color online) 
Average value of the total angular momentum, in units 
of $\hbar$, of the ground state for $N=4$ atoms as a 
function of $\eta$ and $\hbar\Omega_0/E_{\rm R}$. 
The insets concentrate on two different values of 
$\hbar\Omega_0/E_{\rm R}=40$, and 100, respectively.}
\label{fig:summary1}
\end{center}
\end{figure}

\begin{figure}[t]
\begin{center}
\includegraphics*[width=90mm]{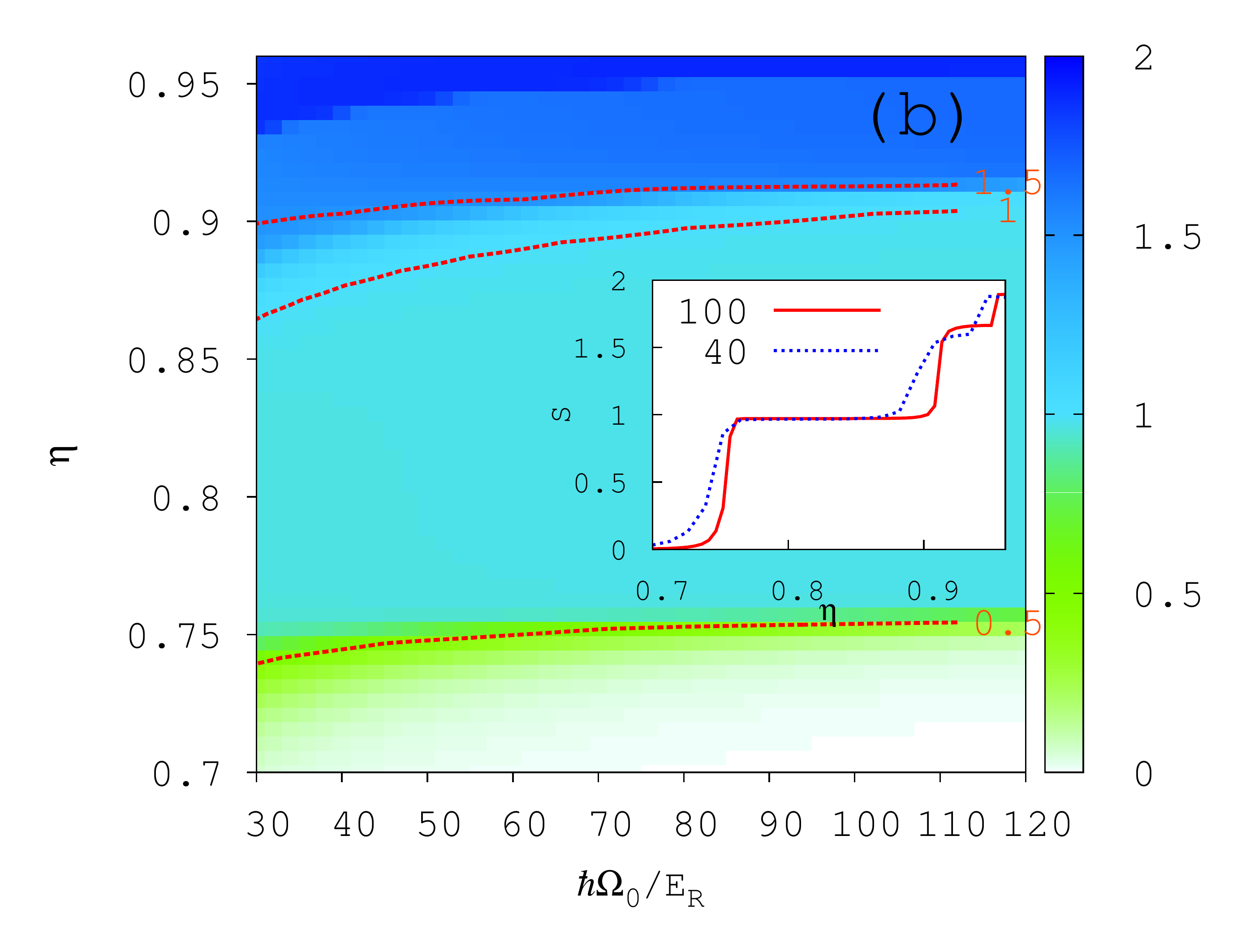}
\caption{(Color online) 
Entropy of the ground state for 
$N=4$ atoms as a function of $\eta$ and $\hbar\Omega_0/E_{\rm R}$. 
The insets concentrate on two different values of 
$\hbar\Omega_0/E_{\rm R}=40$, and 100, respectively.}
\label{fig:summary2}
\end{center}
\end{figure}

In Fig.~\ref{fig:summary1} we show the expectation value of the 
total angular momentum of the GS as a function of 
$\eta$ and $\hbar\Omega_0/E_{\rm R}$. For large $\Omega_0$ we recover the 
step-like structure that is well known for rotating bosonic gases, 
with plateaus at $L=0, 4, 8$, and 12~\cite{Barberan:2006,Cooper:2008} 
with $\eta \omega_\perp$ playing the role of the 
rotating frequency. For an axisymmetric potential containing 
$N\gg 1$ bosons it is well known that the value $L=N$ corresponds 
to a single centered vortex, described by the mean-field state 
$
\Psi_{1\rm vx}= \prod_{i=1}^{N} z_i e^{-\sum z_i^2/2\lambda_{\perp}}\,.
$
Here the squared overlap between our GS and $\Psi_{1\rm vx}$ is 
relatively low (0.47). This is due to the small value of $N$ 
which causes significant deviations from the mean-field 
prediction. The value $L=N(N-1)$ (here $L=12$) in the axisymmetric 
case corresponds to the exact Laughlin state, with a filling 
factor $1/2$ for any $N$. For decreasing values of $\Omega_0$ 
the transitions between the plateaus become broader and are 
displaced towards smaller values of $\eta$. The 
Laughlin-like region is defined here as the interval of 
$\eta$ fulfilling $\langle {\rm GS}| \hat{L} | {\rm GS}\rangle >N(N+1)$.

\subsection{Entropy}
\label{ss2}

An interesting measure for the correlations in the ground state  
is provided by the one-body entanglement entropy~\cite{Horodecki:2009} 
defined as,  
\beq
S=-\mbox{Tr}\left[\rho^{(1)}\ln \rho^{(1)} \right]\,.
\label{entro}
\eeq
Here $\rho^{(1)}$ is the one-body density matrix associated to the 
GS wave-function defined as, 
 \begin{equation}
\rho^{(1)}(\bs r,\bs r') 
= \langle {\rm GS}| \hat{\Psi}^{\dagger}(\bs r) 
\hat{\Psi}(\bs r') |{\rm GS}\rangle
\end{equation}
where $\hat{\Psi}(\vec{r})$ is the field operator, 
$\hat{\Psi}(\vec{r})=\sum_{\ell}\phi_{\ell,0}\hat{a}_{\ell,0}$, 
with $\hat{a}_{\ell,0}$ the operator that destroys a particle with 
angular momentum $\ell$ in the LLL. The natural orbitals, 
$\phi_i({\bs r})$, and their corresponding occupations, $n_i$, 
are defined by the eigenvalue problem, 
\begin{equation}
\int d {\bs r} \rho^{(1)} ({\bs r},{\bs r'}) 
\phi_i({\bs r})=n_i \phi_i(\bs r')\,.
\end{equation}

The entropy of Eq.~(\ref{entro}) provides 
information of the degree of condensation/fragmentation 
of the system and of the entanglement between one 
particle and the rest of the system. This 
entropy definition is enough for our characterization 
of the strong correlations of the ground state. More 
detailed studies, such as whether our ground state 
fulfills area laws for the entanglement entropy~\cite{eisert} 
are beyond the scope of the present paper. The entropy can 
be explicitly evaluated as $S=-\sum_i n_i \ln n_i$. 
Thus it can be checked that this entropy is zero for a true 
Bose--Einstein condensate, since all particles occupy the 
same mode ($n_1=1, n_i=0, i>1$). As the system looses condensation, with 
more than one non-zero eigenvalue, $S$ increases. For the Laughlin 
wave-function with $N$ bosons, $2N-1$ single-particle states are 
approximately equally 
populated, and the entropy is $\sim \ln(2N-1)$.
The entropy $S$ is plotted in Fig.~\ref{fig:summary2} and it 
presents features that are similar to that of Fig.~\ref{fig:summary1}. 
For a fixed $\eta$, the entropy decreases 
with $\Omega_0$. For fixed $\Omega_0$ the dependence on $\eta$ 
exhibits steps, similarly to that of $\langle L\rangle$.  
The region of $\langle L\rangle=0$ corresponds to a fairly condensed 
region with $S\sim 0$. In the one vortex region, corresponding to 
$\langle L\rangle=N$, the 
condensation is already not complete. This is reflected in the 
abovementioned low value of the squared overlap between the 
GS and $\Psi_{1\rm vx}$ as well as in the entropy close to 1. 
Finally, it gradually increases as we increase $\eta$, and reaches 
its maximum value in the Laughlin-like region, $\eta>0.93$.

\subsection{Interaction energy}
\label{ss3}

\begin{figure}[t]
\begin{center}
\includegraphics*[width=90mm]{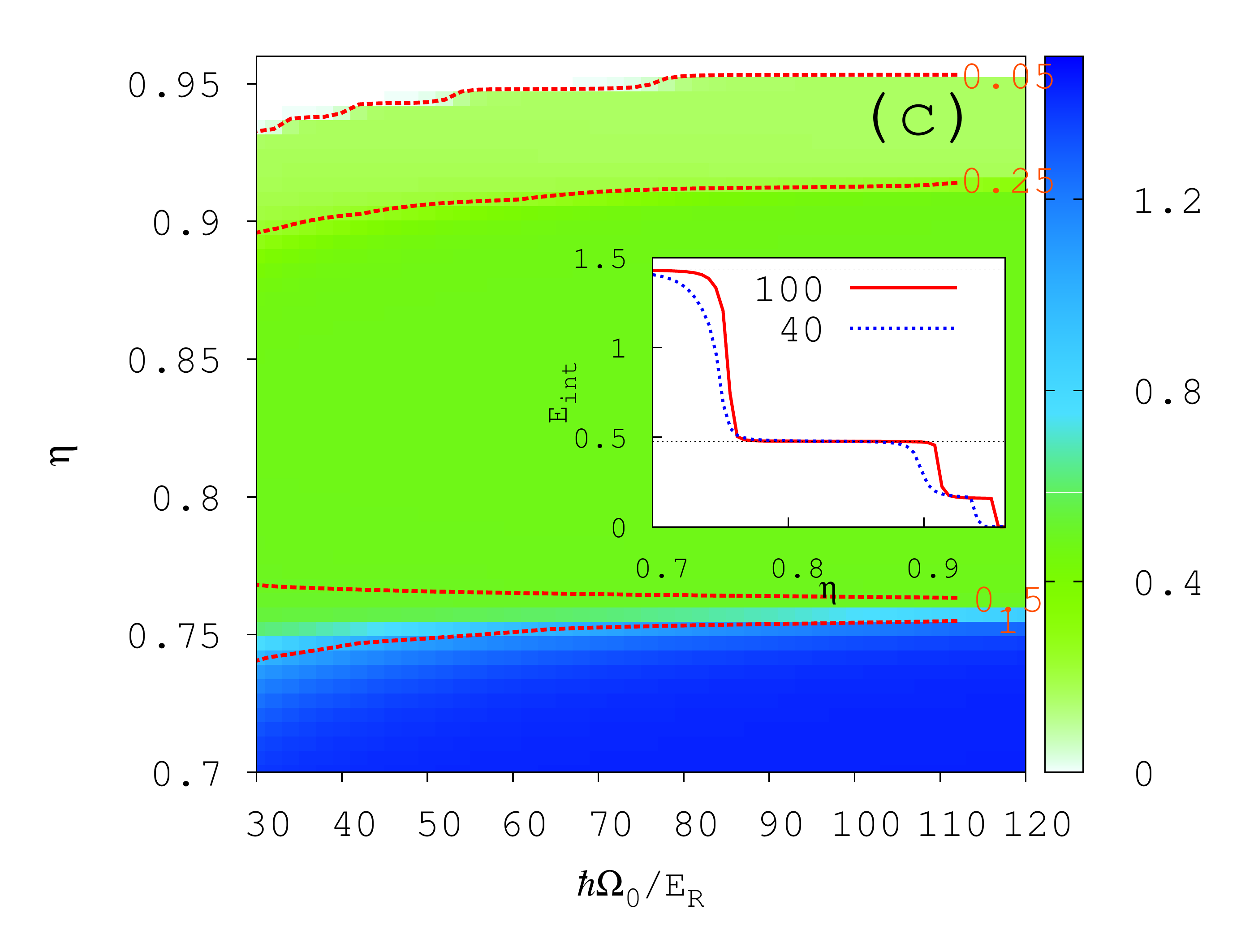}
\caption{(Color online) 
Interaction energy, in units of $\hbar \omega_\perp$, 
of the ground state for $N=4$ atoms as a function of 
$\eta$ and $\hbar\Omega_0/E_{\rm R}$. 
The insets concentrate on two different values of 
$\hbar\Omega_0/E_{\rm R}=40$, and 100, respectively.}
\label{fig:summary3}
\end{center}
\end{figure}

In Fig.~\ref{fig:summary3} we depict the average interaction energy 
as a function of $\eta$ and $\hbar\Omega_0/E_{\rm R}$.  In the inset 
we also plot  for $L=0$ and $L=N=4$ the analytical result expected 
in an axisymmetric potential, $E_{\rm int} = gN (2N - L - 2) / (8 \pi )$, 
valid for $L=0$ and $2\leq L\leq N$  \cite{Bertsch:1999}. The 
interaction energy approaches zero as we increase $\eta$, indicating 
the Laughlin-like nature of the states in the region $\eta\geq 0.93$.

The standard bosonic Laughlin state (at half filling) has the analytical 
form~\cite{laughlin,Cooper:2001,Regnault:2003}
\begin{equation}
\Psi_{\cal L}(z_1,\dots,z_N) ={\cal{N}}\prod_{i<j}(z_i-z_j)^2{\mathrm e}^{-\sum \mid z_i\mid^2/ 2
\lambda_{\perp}^2}\, ,
\label{laughwave}
\end{equation}
where $\cal{N}$ is a normalization constant and $z=x+iy$. 
It is the exact ground state of the system for the 
contact interaction in the adiabatic case\cite{wil}. The contribution 
of the interaction to the energy of the system is zero due to the 
zero probability to have two particles at the same place.

\subsection{Internal correlations}
\label{ss4}

\begin{figure}[t]
\begin{center}
\includegraphics*[width=82mm,angle=-90]{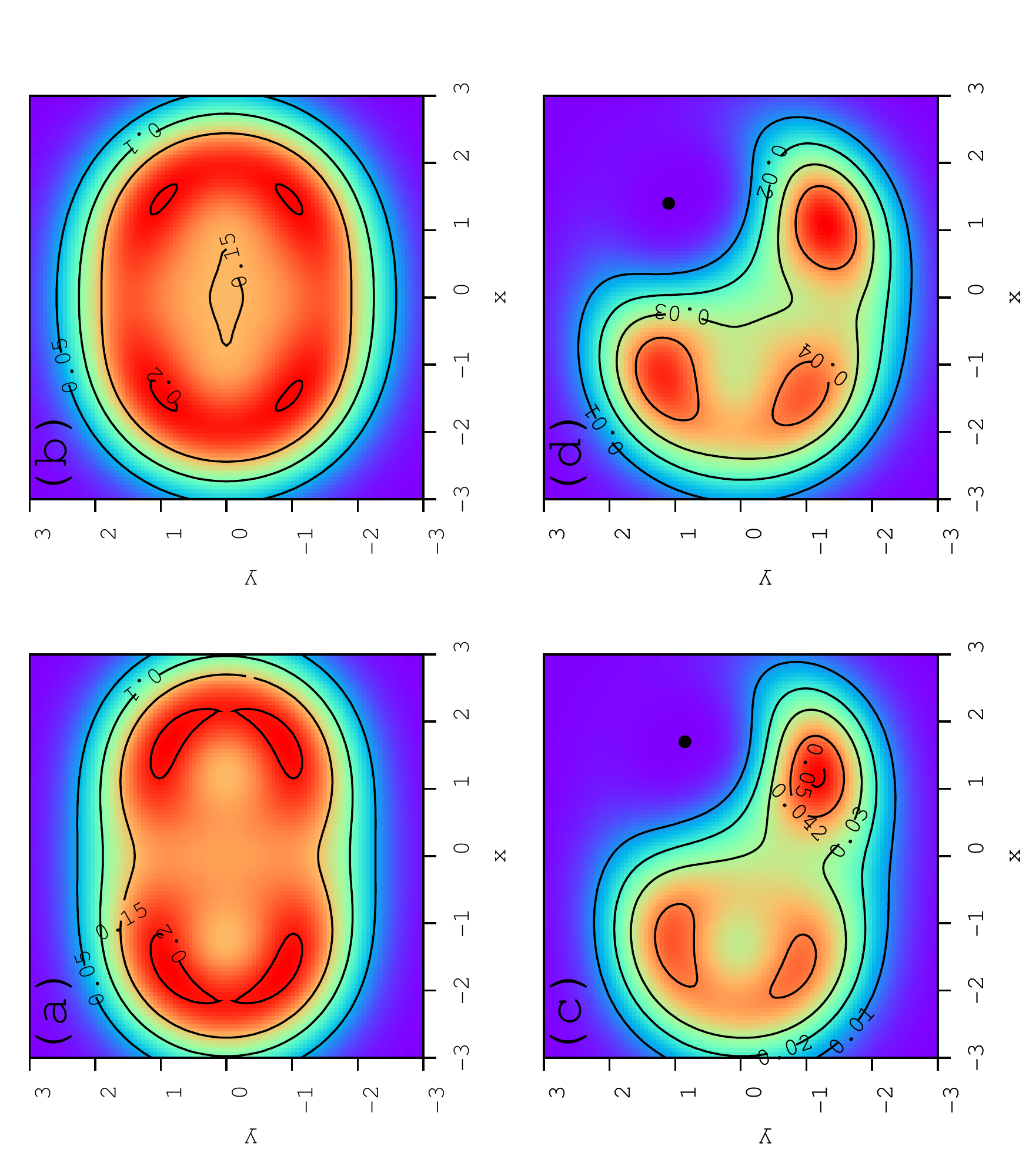}
\caption{(Color online) Density of atoms, panels (a) and (b), and 
pair correlation computed as explained in the text, panels (c) 
and (d), of the ground state for $\hbar\Omega_0/E_R=40$ (a,c) 
and 100 (b,d), respectively. The values of $\eta$ used are 
0.942 and 0.955 for $\hbar\Omega_0/E_R=40$ and 100, respectively.
The solid circle marks the position of $\vec{r}_{\rm 0}$. The length unit 
is $\lambda_\perp$.
\label{fig6}}
\end{center}
\end{figure}

The pair correlation function provides a test for the 
presence of spatial correlations in a system. For the GS 
it is defined as,
\begin{equation}
{\rho}^{(2)}(\vec{r},\vec{r}_0)=
\left\langle {\rm GS} \left|\,\hat{\Psi}^{\dag}(\vec{r}) 
\hat{\Psi}^{\dag}(\vec{r}_0)\hat{\Psi}(\vec{r}_0)\hat{\Psi}(\vec{r})\,
\right|{\rm GS}\,\right\rangle \,.
\end{equation}
In Fig.~\ref{fig6}, panels (c,d), 
$\rho^{(2)}(\vec{r},\vec{r}_{\rm M})$ is depicted, where 
$\vec{r}_{\rm 0}$ is taken as the maximum of the 
corresponding density, depicted in panels (a,b). 
As seen in the figure, once one particle 
is detected in $\vec{r}_{\rm 0}$, the other three appear 
localized at the remaining three vertices of a rectangle. 
This feature, present also in the exact Laughlin wave function, 
survives both for $\hbar \Omega_0/E_R=40$ and 100, even 
though the squared overlap of the ground state with the 
Laughlin differs almost by a factor two, as will be discussed 
later in Sect.~\ref{ss6}. In the adiabatic case, this spatial 
correlation could be inferred from the structure of the 
analytical expression, i.e. the particles tend to avoid each 
other to minimize energy. This is responsible for the particular 
spatial correlation shown in panels (c) and (d).  No other 
ground state with $L<N(N-1)$ exhibits this property. A 
similar phenomenology was found for fermions~\cite{oster}.

\subsection{Energy spectrum}
\label{ss5}

\begin{figure}[t]
\includegraphics*[width=83mm]{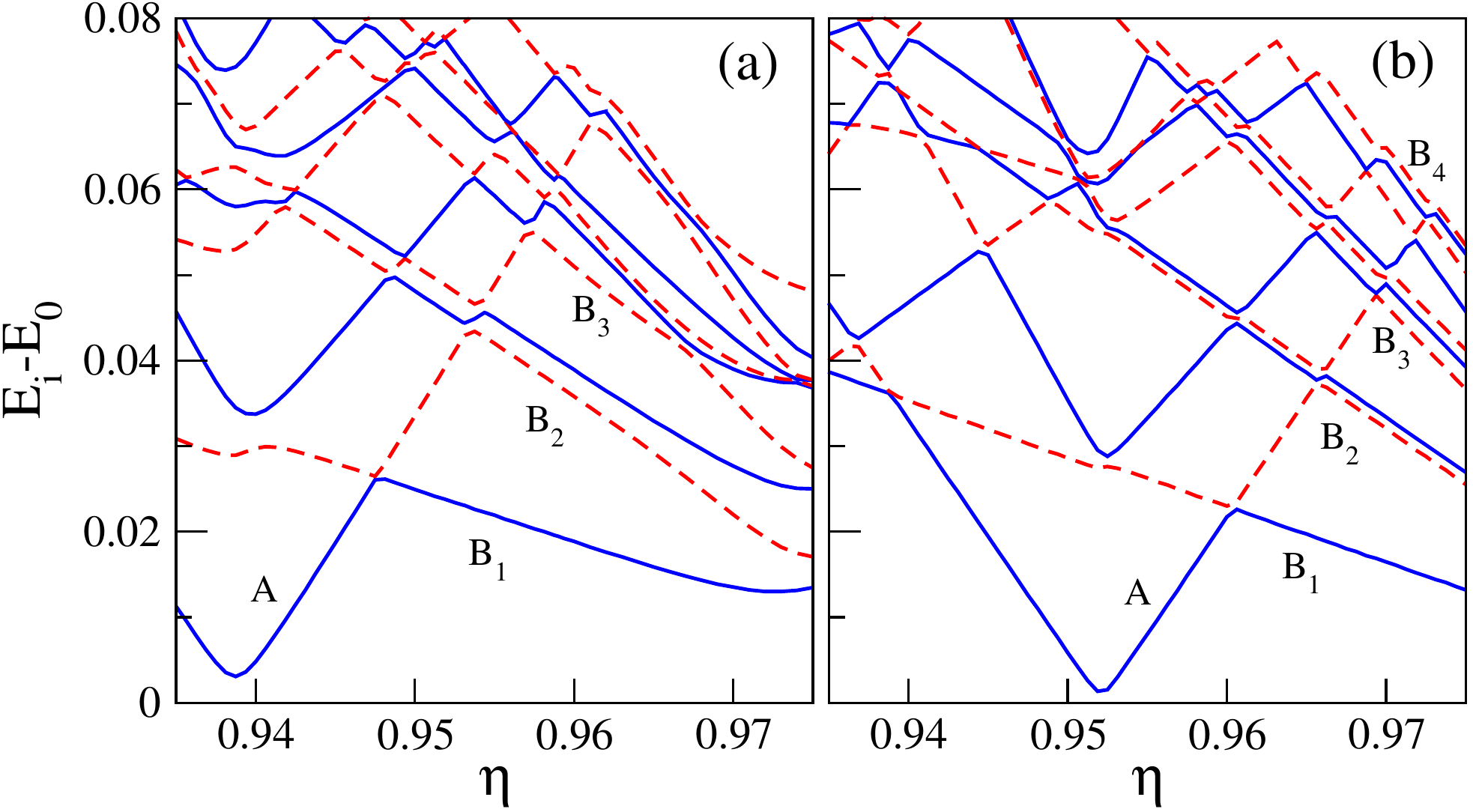}
\caption{(Color online) 
Energy difference in units of $g\,\hbar \omega_\perp$ 
between the first 10 levels of the spectrum and 
the ground state energy as function of $\eta$ for 
$\hbar\Omega_0/E_{\rm R}=40$ (a) and 100 (b). 
\label{fig7}}
\end{figure}

To further characterize the properties of the system we discuss 
the properties of the low energy spectrum, and its evolution 
as we decrease $\Omega_0$, i.e. increasing the non-adiabaticity. 
In Fig.~\ref{fig7} we show the energy difference between 
the ground state and the first ten 
excitations as a function of $\eta$. First, let us recall that 
in the adiabatic case the spectrum of the system has already 
been studied in the context of rotating atomic 
clouds~\cite{Cooper:2008}, thus our main interest here will be 
to characterize the non-adiabatic effects. 

Let us first consider the most symmetric case, Fig.~\ref{fig7} (b).  
In the Laughlin region ($\eta>0.952$) there are two types of 
lowest excitations: quasi-particle and edge excitations. 
For $·0.952<\eta<0.961$ the excitation with $L=N(N-1)-N$ 
marked as $A$ is a quasi-particle type state, while for 
$\eta>0.961$ the state with $L=N(N-1)+1$ marked as $B_1$ is 
an edge excitation. The tower of edge excitations of the 
system are marked as $B_n$ in the figure and 
correspond in the adiabatic case to excitations with 
$L=N(N+1)+n$, with $n>0$. They are fully degenerate in 
the adiabatic case with a degeneracy given by the partition 
function of $n$,  $p(n)$, defined as the number of distinct 
ways in which $n$ can be written as a sum of smaller 
non-negative integers, i.e. $5$ if $n=4$ ~\cite{caza}. 
In panel (b) the degeneracy is partly lifted due to the 
slight non-adiabaticity and in panel (a) the condition 
is clearly relaxed. This structure of the edge 
excitations is a fingerprint of the Laughlin state.

Finally, the maximun energy separation between the ground 
state and its first excitation in the Laughlin-region, 
which in our confined case is both a quasi-particle 
excitation and an edge excitation, increases when 
decreasing $\Omega_0$. It changes  
from $\sim 0.022 g\, \hbar\omega_{\perp}$ for 
$\Omega_0/E_R=100$ to $\sim 0.027 g\, \hbar\omega_{\perp}$ for 
$\Omega_0/E_R=40$. The bulk energy difference in the 
non-adiabatic cases can be estimated by linearly extrapolating 
the segment $A$ to $\eta=1$, giving 
$\sim 0.18 g \,\hbar\omega_{\perp}$ and 
$\sim 0.13 g \,\hbar\omega_{\perp}$ for $\Omega_0/E_R=40$ and 100,
respectively. Thus, by increasing the laser intensity, the bulk 
energy difference approaches the value of the gap reported in 
Ref.~\cite{Regnault:2003} for a symmetric and edgeless system.

\section{Analytical representation of the ground state 
in the Laughlin-like region}
\label{ss6}

\begin{figure}[t]
\includegraphics*[width=83mm]{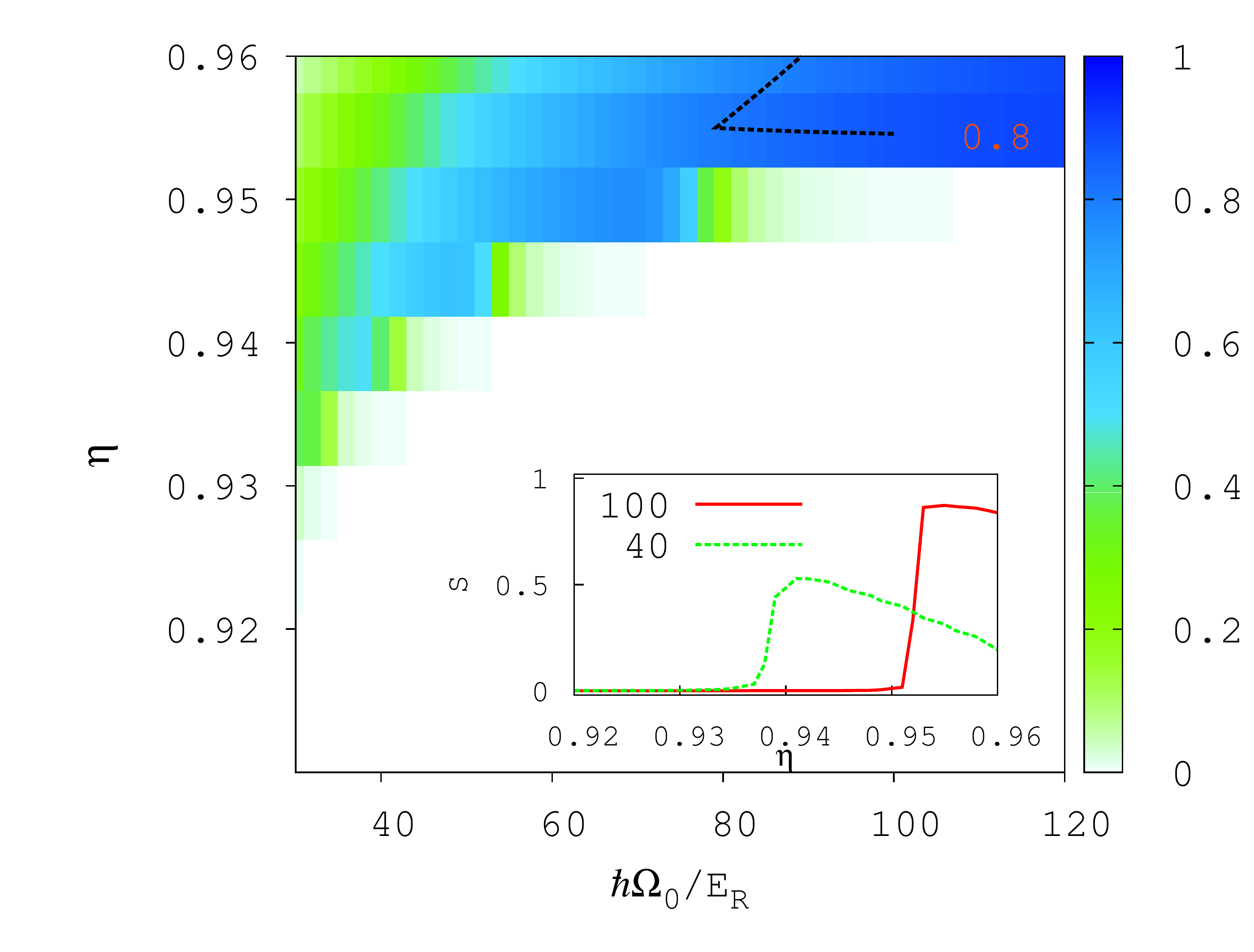}
\caption{(Color online) Squared overlap 
$|\langle \rm GS| \Psi_{\cal L}\rangle|^2$ 
as a function of $\eta$ and $\hbar\Omega_0/E_{\rm R}$ for 
$N=4$. The dashed line marks the region of squared overlap 
larger than 0.8. The inset depicts the squared overlap for 
$\hbar\Omega_0/E_{\rm R}=100$ (solid) and $40$ (dashed) as a 
function of $\eta$.
\label{fig5}}
\end{figure}

In this section, we calculate the overlap of the exact solutions 
for the GS in the Laughlin-like region with several analytical 
expressions. To begin with, we calculate the dependence of the 
squared overlap $|\langle \Psi_{\cal L}| {\rm GS} \rangle|^2$ of 
the Laughlin state with the exact GS as a function of the magnetic 
field strength $\eta$ and the atom--laser coupling $\Omega_0$. The 
result is plotted in Fig.~\ref{fig5}.  For large $\Omega_0$ 
(typically $>80\,E_{\rm R}/\hbar$), the adiabatic approximation 
holds ($H_{22}\approx H_{22}^{\rm eff}$): the overlap between 
the GS and the Laughlin state jumps from a quasi-zero to a 
large ($>0.8$) value when the magnetic field strength $\eta$ 
reaches a threshold value. 

\begin{figure}[t]
\includegraphics*[width=83mm]{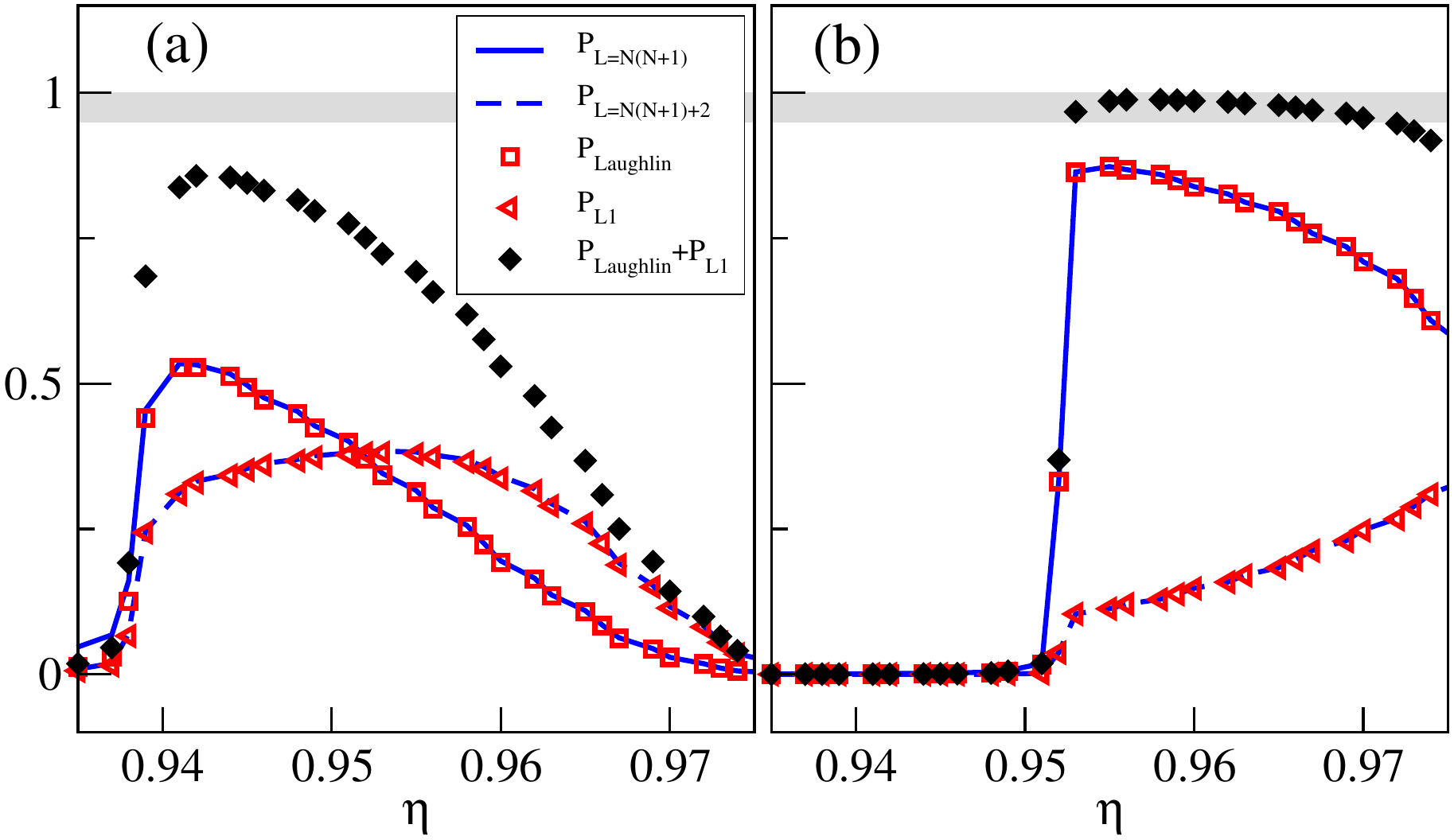}
\caption{(Color online) 
(upper panels) Squared overlaps between the GS of the 
system and the Laughlin wave function, 
$P_{\rm Laug}=|\langle {\rm GS} |\Psi_{\cal L}\rangle|^2$ (solid squares) 
and $\Psi_{{\cal L}1}$, $P_{L1}=|\langle {\rm GS} |\Psi_{{\cal L}1}\rangle|^2$ 
(triangles). The sum of both is depicted as solid diamonds. The solid 
and dashed lines correspond to the weights $P_{L=N(N-1)}$ and 
$P_{L=N(N-1)+2}$ in the GS, respectively. Panel (a) corresponds 
to $\hbar\Omega_0/E_{\rm R}=40$ and (b) to 100. The shaded band 
marks the region of squared overlap larger than 0.95.
(lower panels) Energy difference in units of $\hbar \omega_\perp$ 
between the first 10 levels of the spectrum and the ground state 
energy as function of $\eta$ for $\hbar\Omega_0/E_{\rm R}=40$ (c) and 100 (d). 
\label{fig8}}
\end{figure}

For smaller $\Omega_0$ the overlap is much smaller even for 
large $\eta$ (upper left corner of Fig.~\ref{fig5}). In this 
case the GS must have Jastrow factors that bring the angular momentum 
around the value $L=N(N-1)$ and that suppress interactions 
[Figs.~\ref{fig:summary1} and~\ref{fig:summary3}]. Based on 
these observations, we propose an analytical ansatz for this 
GS of the form
\beq \Psi_{\cal GL} =\alpha \Psi_{\cal L} + \beta \Psi_{{\cal L}1} 
+ \gamma \Psi_{{\cal L}2} \, ,
\label{newl} 
\eeq
with 
\bea
\Psi_{{\cal L}1} &=& {\cal N}_1\, \Psi_{\cal L} \cdot \sum_{i=1}^N
z_i^2\nonumber \\
\Psi_{{\cal L}2}&=& {\cal N}_2\, 
\big(\tilde{\Psi}_{{\cal L}2}- \langle
\Psi_{{\cal L}1}|\tilde{\Psi}_{{\cal L}2}\rangle \Psi_{{\cal
    L}1}\big)\nonumber \\
\tilde{\Psi}_{{\cal L}2}&=& \tilde{{\cal N}}_2 \,\Psi_{\cal L} \cdot
\sum_{i<j}^N z_iz_j \,,
\eea
such that we ensure 
$\langle \Psi_{\cal  L}|\Psi_{{\cal L}i}\rangle=0$ and 
$\langle \Psi_{{\cal L}i}|\Psi_{{\cal L}j}\rangle=\delta_{ij}$. This ansatz 
involves components of angular momentum $L=N(N-1)$ and $L=N(N-1)+2$, and 
zero interaction energy. The coefficients $\alpha$, $\beta$ and $\gamma$ 
are given by the projections of the exact GS onto $\Phi_{\cal L}$, $\Phi_{{\cal L}1}$ 
and $\Phi_{{\cal L}2}$ respectively. In Fig.~\ref{fig8} (a,b) we present the squared 
overlaps $P_{\rm Laughlin}$ and $P_{L_1}$ between the exact GS wave function and the 
functions $\Psi_{\cal L}$ and $\Psi_{{\cal L}1}$, respectively. We 
restrict our study to the Laughlin-region. We also plot the 
weights of the angular momentum subspaces in the GS, $P_{L=N(N-1)}$ 
and $P_{L=N(N-1)+2}$.

The first result of our numerical analysis is that $P_{{\cal L}_2}$ 
is negligible ($<0.005$) over the whole range of Fig.~\ref{fig8}. Then, we note
that the relations $P_{L=N(N-1)} \approx P_{\cal L}$ and $P_{L=N(N-1)+2}
\approx P_{{\cal L}1}$ hold over this range. This implies that the deviation
with respect to the adiabatic approximation mostly increases the weight of the
$\Psi_{{\cal L}1}$ component in the GS. For small values of $\hbar\Omega_0/E_{\rm
  R}$, the squared overlap with the proposed ansatz reaches values of $\sim$
0.85, with the weight of $\Psi_{\cal L}$ and $\Psi_{{\cal L}1}$ being of
comparable size. As $\hbar\Omega_0/E_{\rm R}$ increases above $80$, the GS is very
well represented by Eq.~(\ref{laughwave}), as already explained. Considering 
different particle numbers from $N=3$ to $N=5$, we always find a very similar 
behavior and thus conclude that Eq.~(\ref{newl}) quite generally 
provides a good representation of the GS in the Laughlin-like 
region\footnote{Note that the Laughlin-like region 
decreases notably in size as $N$ is increased.}.

\section{Summary and conclusions}
\label{s4}

In conclusion we have performed exact diagonalization to analyze the ground
state of a small cloud of bosonic atoms subjected to an artificial gauge
field. Our approach allowed us to explore both the regime of very large
atom--laser coupling, where the adiabatic approximation is valid, and the case
of intermediate coupling strengths. In the first case we recovered the known
results for a single component gas in an axisymmetric  potential. The second
case is crucial for practical implementations because it requires less
light intensity on the atoms, which decreases the residual heating due to
photon scattering.  In this case we have identified a regime where a strongly
correlated ground state emerges, which shares many similarities with the
Laughlin state in terms of angular momentum, energy, internal 
spatial correlations and lowest excitations, although the overlap between 
the two remains small. Importantly, a reduction of the laser intensity shifts the region where
Laughlin-like states exist to lower values of the effective 
magnetic field, thus departing from the instability region, $\eta>1$. 
We have also proposed an ansatz 
that represents the ground state quite accurately for a region of the 
parameter space. Finally, let us emphasize that the properties 
analyzed in this article are measurable quantities, as is the case of the 
expected value of the angular momentum, the pair correlation distribution 
and excitation spectrum.

\begin{acknowledgments}
This work has been supported by 
EU (NAMEQUAM, AQUTE, MIDAS), ERC (QUAGATUA), 
Spanish MINCIN (FIS2008-00784, FIS2010-16185, 
FIS2008-01661 and QOIT Consolider-Ingenio 2010), 
Alexander von Humboldt Stiftung, IFRAF and ANR (BOFL project). 
B.~J.-D. is supported by a Grup Consolidat SGR 21-2009-2013. 
\end{acknowledgments}

\vskip1cm
\appendix
\section{Explicit form of the $H_{21}H_{12}$ term}
\label{ap1}

We provide here the explicit expression for the 
term $H_{21}H_{12}$ appearing in the perturbatively derived Hamiltonian 
$H_{22}^{\rm eff}$. As explained in the text we consider 
up to quadratic terms in $x$ and $y$. The explicit expression 
then reads
\bea
H_{21}H_{12} &=& 
\left(\frac{\hbar ^4}{4 M^2 w^4}-\frac{2 x^2 \hbar ^4}{M^2 w^6}+\frac{k^2 x^2
  \hbar ^4}{16 M^2 w^4}+\frac{k^4 x^2 \hbar ^4}{64 M^2 w^2}\right.\nonumber \\
&&\left.+\frac{i k x y
  \hbar ^4}{4 M^2 w^5}+\frac{k^2 y^2 \hbar ^4}{64 M^2 w^4}\right) \nonumber \\
&&+\left(-\frac{i k x \hbar ^4}{4 M^2 w^3}-\frac{i k^3 x \hbar ^4}{8 M^2
  w}\right) \partial_y \nonumber \\
&&+\left(\frac{x \hbar ^4}{M^2 w^4}-\frac{i k y \hbar ^4}{8 M^2 w^3}\right) \partial_x \nonumber \\
&&+\left(-\frac{k^2 \hbar ^4}{4 M^2}+\frac{k^2 x^2 \hbar ^4}{4 M^2 w^2}\right)
\partial^2_y \nonumber \\
&&+\left(-\frac{\hbar ^4}{4 M^2 w^2}+\frac{x^2 \hbar ^4}{2 M^2 w^4}\right)
\partial^2_x \,.
\eea

\end{document}